\documentclass[conference]{IEEEtran}
\IEEEoverridecommandlockouts
\usepackage{cite}
\usepackage{amsmath,amssymb,amsfonts}
\usepackage{algorithmic}
\usepackage{graphicx}
\usepackage{textcomp}
\usepackage{hyperref}
\hypersetup{hidelinks}
\usepackage{svg}
\makeatletter
\let\MYcaption\@makecaption
\makeatother
\usepackage[font=footnotesize, skip=0pt]{caption}
\usepackage[font=footnotesize, skip=0pt]{subcaption}
\makeatletter
\let\@makecaption\MYcaption
\makeatother
\usepackage{xcolor}
\usepackage{color, colortbl}
\usepackage{listings}
\usepackage{multirow}
\usepackage{hhline}
\usepackage{array}



\newif\ifarxiv

\arxivtrue

\ifarxiv
\makeatletter
\newcommand*\titleheader[1]{\gdef\@titleheader{#1}}
\AtBeginDocument{%
  \let\st@red@title\@title%
  \def\@title{%
    \bgroup\normalfont\normalsize\centering\@titleheader\par\egroup
    \vskip0.5em\st@red@title}
}
\makeatother
\fi

\def\BibTeX{{\rm B\kern-.05em{\sc i\kern-.025em b}\kern-.08em
    T\kern-.1667em\lower.7ex\hbox{E}\kern-.125emX}}

\newcommand\ceil[1]{\lceil#1\rceil}
\ifarxiv
\newcommand{\coloredtxt}[1]{\textcolor{black}{#1}}
\else
\newcommand{\coloredtxt}[1]{\textcolor{black}{#1}}
\fi
\newcommand\theacronym{MCCM}

\definecolor{seg_color}{HTML}{1f77b4}
\definecolor{segrr_color}{HTML}{ff7f0e}
\definecolor{hyb_color}{HTML}{2ca02c}

\title{An Analytical Cost Model for Fast Evaluation of Multiple Compute-Engine CNN Accelerators}

\ifarxiv
\titleheader{Accepted for publication in 2025 IEEE International Symposium on Performance Analysis of Systems and Software (ISPASS)}
\fi

\begin{document}

\author{
\IEEEauthorblockN{Fareed Qararyah \href{https://orcid.org/0000-0002-3955-2836}{\includegraphics[width=1em]{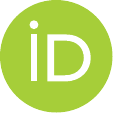}}, Mohammad Ali Maleki \href{https://orcid.org/0000-0002-9019-3605}{\includegraphics[width=1em]{figures/orcid_logo.pdf}}, Pedro Trancoso \href{https://orcid.org/0000-0002-2776-9253}{\includegraphics[width=1em]{figures/orcid_logo.pdf}}}
\IEEEauthorblockA{\textit{Department of Computer Science and Engineering} \\
\textit{Chalmers University of Technology and University of Gothenburg, Gothenburg, Sweden}\\
\{qarayah, mohammad.ali.maleki, ppedro\}@chalmers.se}
\ifarxiv
\thanks{\copyright\ 2025 IEEE. Personal use of this material is permitted. Permission from IEEE must be obtained for all other uses, in any current or future media, including reprinting/republishing this material for advertising or promotional purposes, creating new collective works, for resale or redistribution to servers or lists, or reuse of any copyrighted component of this work in other works.}
\fi
}

\maketitle

\thispagestyle{plain}
\pagestyle{plain}

\begin{abstract}
Convolutional Neural Networks (CNNs) serve various applications with diverse performance and resource requirements. Model-aware CNN accelerators best address these diverse requirements. These accelerators usually combine multiple dedicated Compute Engines (CEs). The flexibility of Field-Programmable Gate Arrays (FPGAs) enables the design of such multiple Compute-Engine (multiple-CE) accelerators. However, existing multiple-CE accelerators differ in how they arrange their CEs and distribute the FPGA resources and CNN operators among the CEs. The design space of multiple-CE accelerators comprises numerous such arrangements, which makes a systematic identification of the best ones an open challenge.\\
This paper proposes a \underline{M}ultiple-\underline{C}E accelerator analytical \underline{C}ost \underline{M}odel (\theacronym) and an evaluation methodology built around \theacronym. The model and methodology streamline the expression of any multiple-CE accelerator and provide a fast evaluation of its performance and efficiency. \theacronym~is in the order of $100000 \times$ faster than traditional synthesis-based evaluation and has an average accuracy of $> 90\%$. The paper presents three use cases of \theacronym. The first describes an end-to-end evaluation of state-of-the-art multiple-CE accelerators considering various metrics, CNN models, and resource budgets. The second describes fine-grained evaluation that helps identify performance bottlenecks of multiple-CE accelerators. The third demonstrates that \theacronym~fast evaluation enables exploring the vast design space of multiple-CE accelerators. These use cases show that no unique CE arrangement achieves the best results given different metrics, CNN models, and resource budgets. They also show that fast evaluation enables design space exploration, resulting in accelerator designs that outperform state-of-the-art ones. \coloredtxt{\href{https://github.com/fqararyah/MCCM}{\theacronym~is available at https://github.com/fqararyah/MCCM}}.
\end{abstract}

\begin{IEEEkeywords}
Cost model, multiple engine accelerators, field-programmable gate arrays (FPGAs), convolutional neural networks, inter-layer pipelining.
\end{IEEEkeywords}

\section{Introduction}

The use of Convolutional Neural Networks (CNNs) in various application domains and across a spectrum of environments, ranging from edge to cloud, brings a wide range of performance requirements and optimization goals~\cite{venieris2018fpgaconvnet, choudhury2022fpga}. While some applications are latency-critical, others target high throughput. Moreover, resource efficiency is crucial when having limited resources~\cite{chung2018serving,xia2021sparknoc}, and when resources are shared by multiple applications~\cite{venieris2023multiple}. FPGAs' reconfigurability enables the design of accelerators tailored to such diverse requirements, which makes FPGA-based CNN accelerators popular~\cite{fowers2018configurable, chung2018serving, zhang2018dnnbuilder, blott2018finn, wu2019high, xia2021sparknoc}. 

Accelerators that organize FPGA resources into generic reusable Compute Engines (CEs) have limited adaptability to the varying characteristics of CNN layers, leading to dynamic resource underutilization~\cite {wei2018tgpa, zhang2020dnnexplorer,qararyah2022fibha,cai2022deepburning, qararyah2024efficient}. This limitation is often described as one size does not fit all. Multiple Compute-Engine (\emph{multiple-CE}) accelerators overcome this limitation. They organize FPGA resources into a number of dedicated CEs tailored to the structure of the CNN model and the available resources~\cite{alwani2016fused, zhang2020dnnexplorer, nguyen2020layer, qararyah2024efficient, shen2016overcoming, shen2017maximizing, venieris2017latency, wei2018tgpa, cai2022deepburning}. However, the existing multiple-CE accelerators differ in how they arrange their CEs and distribute the FPGA resources and CNN layers among the CEs. Identifying CE arrangements that achieve the best performance and efficiency requires a systematic approach to multiple-CE accelerator design.

A systematic accelerator design approach requires analyzing the \emph{bottlenecks} and inefficiencies and \emph{exploring the design space} to identify optimizations that mitigate them. The literature reveals that the inefficiencies of Deep Learning (DL) accelerators, in general, have three root causes, \textbf{(1)} \emph{Processing Element (PE) underutilization}~\cite{boroumand2021google, ma2018optimizing, zhang2022full}, \textbf{(2)} \emph{large on-chip buffers}~\cite{boroumand2021google, petrica2020memory, yu2024auto}, \textbf{(3)} and the time and energy costly \emph{off-chip access}~\cite{zhang2022full, ma2018optimizing}. Considering multiple-CE accelerators, the second and third causes are still open challenges. However, multiple-CE accelerators have less PE underutilization than generic ones. Multiple-CE accelerators mitigate PE underutilization bottleneck but in a way that creates \emph{throughput and latency} trade-off~\cite{wei2018tgpa, qararyah2024efficient, cai2022deepburning}. Hence, a systematic multiple-CE accelerator design approach requires analyzing PE underutilization with latency-throughput trade-offs, on-chip buffer requirements, and off-chip access. 

The large design space of CE arrangement possibilities and the lack of a fast methodology that models and evaluates their impact on multiple-CE accelerators' performance and efficiency make a systematic design approach impractical. While High-Level Synthesis (HLS) shortens the design time, synthesizing a single accelerator instance can take hours. To overcome this challenge, this paper proposes a Multiple-CE accelerator analytical Cost Model (\theacronym), and an evaluation methodology built around it. The proposed model and methodology enable the expression of any multiple-CE accelerator using simple notation and provide a fast and accurate evaluation of its throughput, latency, on-chip buffer requirements, and off-chip accesses. 
This paper makes the following contributions:

\begin{itemize}
    \item It proposes a multiple-CE accelerator analytical cost model (\theacronym). \theacronym~applies bottom-up modeling of the impact of CE arrangement possibilities on the performance and efficiency of multiple-CE accelerators.
    \item It proposes \theacronym-based evaluation methodology, which enables expressing any multiple-CE accelerator and evaluates its latency, throughput, on-chip buffer requirements, and off-chip accesses orders of magnitude faster than the traditional approach.
    \item It presents three use cases of \theacronym. First, an end-to-end evaluation of state-of-the-art multiple-CE accelerators considering different metrics, CNNs, and resource budgets. Second, fine-grained evaluation of multiple-CE accelerators' performance bottlenecks. Third, a design space exploration of multiple-CE accelerators enabled by \theacronym~fast evaluation.
\end{itemize}

\theacronym~is in the order of $100000 \times$ faster than synthesis-based evaluation and has an average accuracy of $> 90\%$. The presented \theacronym~use cases using three state-of-the-art multiple-CE accelerators, five CNNs, and four FPGA boards show that no single multiple-CE architecture is the best, given different metrics, CNN models, and resource budgets. Moreover, the evaluation shows the impact of engine arrangements on multiple-CE accelerators' performance bottlenecks. Finally, using \theacronym~fast evaluation as a basis for multiple-CE accelerators design space exploration enables identifying designs that outperform the state-of-the-art.

\section{Background and Motivation}
\label{sec:background}

\subsection{Convolutional Neural Networks (CNNs)}

Convolutional Neural Networks (CNNs) are feed-forward Deep learning (DL) algorithms~\cite{lecun2015deep}. A CNN comprises a sequence of \emph{layers} that perform feature extraction and classification~\cite{lecun2010convolutional}.
Convolutional layers are the primary layers in a CNN. A convolutional layer consists of a collection of trainable parameters, known as \emph{weights}, which are organized into \emph{filters}. These filters extract features from multi-dimensional inputs or intermediate results. The inputs of a layer are referred to as Input Feature Maps (\emph{IFMs}), and the outputs as Output Feature Maps (\emph{OFMs}). The term Feature Maps (\emph{FMs}), or activations, refers to both IFMs and OFMs. FMs consist of 2D slices known as \emph{channels}.

\subsection{CNN Compute-Engines}
\label{subsec:ce_parallelism}

\begin{figure}[!htbp]
%\vspace{-10pt}
     \centering
     \captionsetup{justification=centering}
     \includegraphics[width=\columnwidth]{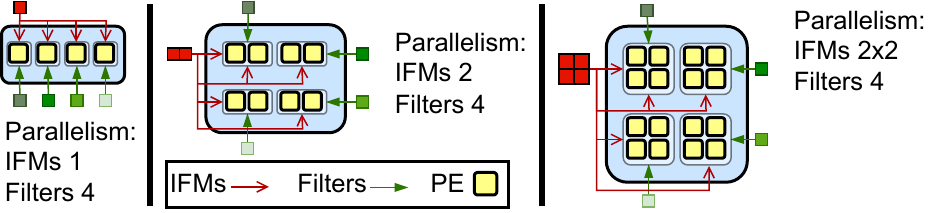}
     \caption{Compute-Engine (CE) parallelism examples}
     %\vspace{-3pt}
     \label{fig:ce_parallelism}
 \end{figure}
 
As convolutions represent more than 90\% of modern CNN operations~\cite{ma2018optimizing}, CNN Compute-Engine (CE) optimizations focus on convolutional layers. A CE comprises a grid of Processing Elements (PEs) where each PE performs a MAC operation, e.g., a DSP in an FPGA. Two main design decisions that affect the performance and efficiency of a CE are \emph{parallelism strategy} and \emph{dataflow}. A convolution operation comprises a loop nest of six loops without batching~\cite{kao2020gamma}. A CE parallelism strategy describes which among these six loops are partially or fully parallelized. The CE parallelism strategy affects data reuse and memory access patterns~\cite{chen2019eyeriss, ma2018optimizing}. Figure~\ref{fig:ce_parallelism} depicts examples of parallelism strategies in 3-D, 2-D, and 1-D. An exhaustive analysis of parallelism strategies on FPGAs shows that parallelizing on three dimensions, namely across filters and within an IFM channel width and height, yields the best average reuse and CE utilization across the layers of common CNNs~\cite{ma2018optimizing}. However, 2-D and 1-D parallelism are also common when each CE has a limited PEs budget or when 1-D or 2-D options are more compatible with the dimensions of the CNN layers processed by that CE~\cite{zhang2015optimizing, ma2016scalable, alwani2016fused, shen2017maximizing, zhang2018dnnbuilder, venieris2016fpgaconvnet, zhang2020dnnexplorer}. In the context of convolution CEs, the term dataflow refers to both convolution loop transformations, \emph{i.e.} the scheduling of the loop computations, and the mapping of computations across the CE PEs. The most common dataflow types are \emph{weight-stationary}, \emph{output-stationary}, and \emph{input-stationary}. Each dataflow indicates which of the IFMs, OFMs, or weights is scheduled to move least frequently~\cite{kwon2019understanding}.

\subsection{Multiple compute-engine CNN accelerators}
\label{subsec:back_multi_ce}

\begin{figure}[!htbp]
%\vspace{-10pt}
     \centering
     \captionsetup{justification=centering}
     \includegraphics[width=\columnwidth]{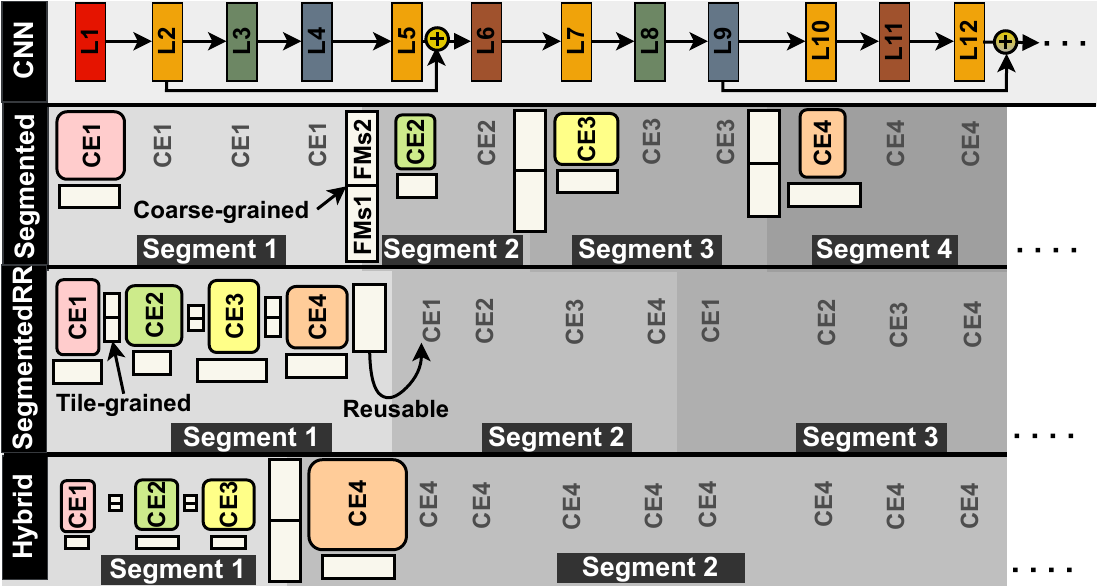}
     \caption{CNN to multiple-CE architecture mapping examples. For example, in Segmented, CE1 processes layers L1-L4, CE2 processes layers L5 and L6, and so on. In practice, CE and buffer sizes are proportional to the segment layers' compute and memory requirements, respectively.}
     \label{fig:multi_ce_mappings_detailed}
     %\vspace{-3pt}
 \end{figure}
 
We use the term multiple Compute-Engine (\emph{multiple-CE}) accelerators to refer to accelerators that organize the FPGA resources into more than one convolution CE. The extreme case of multiple-CE architecture is to have a CE per CNN layer tailored to that layer's characteristics. This approach is resource-demanding and not scalable~\cite{cai2022deepburning, zhang2020dnnexplorer, qararyah2024efficient}. To be both CNN model-aware and scalable, state-of-the-art multiple-CE accelerators adjust the number of CEs based on the CNN characteristics and the available FPGA resources. Exploring the literature shows that, apart from the mentioned extreme case, multiple-CE accelerators follow one of three architectural patterns. Figure~\ref{fig:multi_ce_mappings_detailed} shows a high-level representation of these architectures using four CEs. In practice, the number of CEs varies depending on the CNN structure and the available resources. Each of these CEs may have a unique number of PEs, parallelism strategy, and dataflow (Section~\ref{subsec:ce_parallelism}).
 
We refer to the first architecture in Figure~\ref{fig:multi_ce_mappings_detailed} as \textbf{Segmented}~\cite{shen2016overcoming, shen2017maximizing}. In this architecture, each CE processes one or more \emph{segments}. Each segment consists of a set of consecutive layers. Each CE has its on-chip buffer that stores the weights and FMs partially or fully, depending on the available on-chip memory. \emph{Coarse-grained pipelining}, where different CEs process different inputs (images) at a certain time step, is applied between the CEs. CE inputs are double-buffered to enable pipelining. We call the second architecture \textbf{SegmentedRR} (round-robin) as CEs in such architecture usually process CNN layers circularly according to their topological order~\cite{venieris2017latency, blott2018finn, wei2018tgpa}. While some apply course-grained pipelining between CEs, Wei \emph{et. al}~\cite{wei2018tgpa} have shown that, for this architecture, \emph{tile-grained pipelining }is more efficient. In addition to the double-buffering between the CEs, each CE has a separate buffer for the weights. We call the third architecture \textbf{Hybrid}~\cite{alwani2016fused, zhang2020dnnexplorer, nguyen2020layer, qararyah2024efficient}. It comprises two parts. The first contains a set of tile-grained pipelined CEs, each processing a single layer. The second has a larger CE that processes the rest of the layers. If CNN has two types of convolutional layers, the second part could have two sub-CEs~\cite{qararyah2024efficient}. Coarse-grained pipelining is applied between the Hybrid's first and second parts.

\subsection{Multiple-CE accelerator modeling and evaluation}
\label{subsec:motivation}

\begin{table}
\centering
\caption{Comparison of multiple-CE accelerators using ResNet50 on AMD Zynq 7000 SoC ZCU102. The values for a metric are normalized to the best in that metric.}
%\resizebox{\columnwidth}{!}{
\begin{tabular}{lccc} \hline
\rowcolor{gray!10}
 & latency & on-chip buffers & off-chip accesses \\ \hline 
\cellcolor{gray!10}SegmentedRR & \cellcolor{hyb_color!20}1.0 & 2.64 & 1.79\\ 
\cellcolor{gray!10}Segmented & 4.7 & \cellcolor{hyb_color!20}1.0 & 1.99\\ 
\cellcolor{gray!10}Hybrid (b)& 1.11 & 1.74 & \cellcolor{hyb_color!20}1.0\\ \hline 
\end{tabular}
%}
\label{tab:motivational_atble}
%\vspace{-10pt}
\end{table}

% SegRR & 1.0 & 0.56 & 2.64 & 1.79\\ \hline 
% HYBRID & 1.11 & 1.0 & 1.78 & 1.0\\ \hline 
% SEGMENTED & 4.7 & 0.86 & 1.0 & 1.99\\ \hline 
% HYBRID & 1.11 & 0.99 & 1.74 & 1.0\\ \hline 

State-of-the-art multiple-CE accelerators adopt one variant of the three mentioned architectures and focus on optimizing over two extremes, namely reusable-CEs based accelerators~\cite{bai2018cnn, ma2018optimizing, ye2020hybriddnn} and accelerators with as many CEs as CNN layers~\cite{zhang2018dnnbuilder,umuroglu2017finn}. Adopting fixed architectures limits the potential of existing multiple-CE accelerators. No single architecture is optimal when considering different metrics. Table~\ref{tab:motivational_atble}, which reports normalized synthesis results of instances of the three architectures, exemplifies that. For example, the SegmentedRR at the first row has the best latency but requires $2.64\times$ the on-chip buffer and $1.79\times$ the off-chip accesses of the best accelerator in each metric. Note that buffer requirements and off-chip memory accesses are treated here as optimization goals. This is because minimizing off-chip accesses, consequently bandwidth requirements, and on-chip buffer requirements are crucial, especially when resources are shared by multiple applications~\cite{venieris2023multiple}, even when neither of them forms a bottleneck. The fact that no single architecture gives the best results considering different metrics, even for the same CNN and FPGA as the table shows, suggests that a fixed architecture is far from optimal. Hence, a more systematic design approach that explores different CE arrangement possibilities is needed to identify the best designs given an application performance metrics, CNN model, and resource budget. However, a traditional evaluation of one CE arrangement can take hours. This makes exploring the numerous CE arrangement possibilities of multiple-CE accelerators, and consequently a systematic design approach, impractical. This motivates the proposal of a fast evaluation methodology for multiple-CE accelerators.

\section{Multiple-CE accelerator evaluation methodology}

\subsection{Overview of the evaluation methodology}
\label{subsec:overview}

\begin{figure}[htbp!]
     \centering
     \captionsetup{justification=centering}
     \includegraphics[width=\columnwidth]{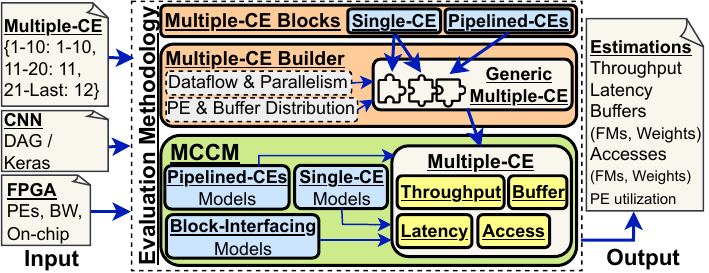}
     \caption{Overview of multiple-CE evaluation methodology.}
     \label{fig:mccm_flow}
     %\vspace{-10pt}
\end{figure}
 
Figure~\ref{fig:mccm_flow} presents an overview of the proposed multiple-CE accelerator evaluation methodology. The methodology takes as \textbf{inputs}: \textbf{(1)} multiple-CE accelerator description, discussed in the Section~\ref{subsec:expressing}, \textbf{(2)} a CNN representation, \textbf{(3)} FPGA's number of PEs (DSPs), off-chip memory bandwidth, and on-chip memory capacity. The main \textbf{outputs} are throughput, latency, on-chip buffer requirements, and off-chip accesses. Other outputs include a fine-grained analysis of PE utilization and a breakdown of the results on the level of weights and FMs. The evaluation methodology comprises three modules. The \textbf{Multiple-CE Blocks} module contains codified descriptions of two building blocks that could be used to construct any multiple-CE accelerator, namely \emph{single-CE} and \emph{pipelined-CEs}. A building block comprises one or more CEs. The \textbf{Multiple-CE Builder} module constructs a \emph{generic multiple-CE accelerator} backbone by combining the building blocks considering the three methodology inputs. It then decides the implementation details, including the CEs buffer and PE distribution, dataflows, and parallelism strategies based on a set of heuristics inspired by the prior art~\cite{shen2017maximizing, blott2018finn, wei2018tgpa, ma2018optimizing, qararyah2024efficient}. The \textbf{\theacronym}~module evaluates a generic multiple-CE accelerator bottom-up by modeling its building blocks and their interfaces. \theacronym~estimates multiple-CE accelerator throughput, latency, on-chip buffers, and memory access requirements. \theacronym~is the main contribution of this work; hence, we use the terms \theacronym~and evaluation methodology interchangeably. \theacronym~components are discussed in detail throughout Section~\ref{sec:mccm}. \coloredtxt{As DL accelerator design is a hot topic where new accelerators are frequently proposed, we designed \theacronym~to be modular and extensible. MCCM interface allows easy addition of models of newly proposed specialized designs.\footnote{The details on how to extend \theacronym~are provided in the repository: {\href{https://github.com/fqararyah/MCCM}{https://github.com/fqararyah/MCCM}}.}}

\subsection{Expressing a generic multiple-CE accelerator}
\label{subsec:expressing}

Examining existing multiple-CE accelerators, shown in Figure~\ref{fig:multi_ce_mappings_detailed}, shows that they comprise two basic building blocks. The first is a \emph{single-CE processing a range of layers} one by one. For example, $CE_1$ processing layers $L_1-L_4$ in the Segmented, and $CE_4$ processing $L_4$ to the last layer in the Hybrid. The second building block is a set of \emph{pipelined-CEs, each processing a layer}. For example, $CE_1-CE_4$ processing layers $L_1-L_4$ in SegmentedRR, and $CE_1-CE_3$ processing layers $L_1-L_3$ in Hybrid. These two blocks could be used to express any multiple-CE architectures. We propose the following notation to express a multiple-CE architecture:
\begin{itemize}
    \item $CE_x$: denotes a single-CE block, and $CE_x-CEy$: denotes $(y - x) + 1$ pipelined-CEs block.
    \item $\{L_x-L_y: CE_z\}$: denotes that layers $x$ to $y$ are processed using a single-CE block ($CE_z$) sequentially. A special case is having one layer only, namely $\{L_x: CE_z\}$.
    \item $\{L_x-L_y: CE_z-CE_w\}$: denotes that layers $x$ to $y$ are processed using pipelined-CEs block. If the number of layers exceeds the number of CEs, the pipelined-CEs block is assumed to process $(w - z) + 1$ layers at a time.
\end{itemize}

% \begin{figure}[!htbp]
%      \centering
%      \captionsetup{justification=centering}
%      \includegraphics[width=\columnwidth]{figures/multi_engine_mappings_illustration.drawio.pdf}
%      \caption{Multiple-CE accelerator building blocks examples}
%      \label{fig:multi_engine_mappings_illustration}
%  \end{figure}
 
For example, the Segmented accelerator in Figure~\ref{fig:multi_ce_mappings_detailed} is expressed as $\{\textit{L}1-\textit{L}4: \textit{CE}1,\ \textit{L}5-\textit{L}6: \textit{CE}2,\ \textit{L}7-\textit{L}9: \textit{CE}3,\ \textit{L}10-\textit{L}12: \textit{CE}4, ... \}$, and the SegmentedRR can be expressed as $\{\textit{L}1-Last: \textit{CE}1-\textit{CE}4\}$. Multiple-CE Builder transforms this notation, given the CNN and FPGA descriptions, into a generic multiple-CE accelerator representation that is fed into the analytical cost model. 

\section{Multiple-CE accelerator analytical cost model}
\label{sec:mccm}

The \underline{M}ultiple-\underline{C}E accelerator analytical \underline{C}ost \underline{M}odel (\theacronym) models a generic multiple-CE accelerator using a bottom-up approach. It evaluates the accelerator using the models of the single-CE and pipelined-CEs blocks and the interfaces between them. The blocks modeling is discussed in Sections~\ref{sec:blocks_modeling}. Modeling a whole accelerator depends on \textbf{(1)} whether each of its blocks processes multiple segments (like the pipelined-CEs block in the SegmentedRR in Figure~\ref {fig:multi_ce_mappings_detailed}) or one segment (like single-CEs blocks in the Segmented in Figure~\ref {fig:multi_ce_mappings_detailed}) and \textbf{(2)} whether there is an inter-segment pipelining (like in the Segmented in Figure~\ref {fig:multi_ce_mappings_detailed}) or not. The details of such dependencies are discussed in Section~\ref{sub_sec:from_b_to_m}.

\subsection{Modeling multiple-CE accelerator building blocks}
\label{sec:blocks_modeling}

\begin{figure*}[htbp!]
 \centering
 \begin{subfigure}[t]{0.36\textwidth}
         \centering
         \captionsetup{justification=centering}
         \includegraphics[width=\textwidth]{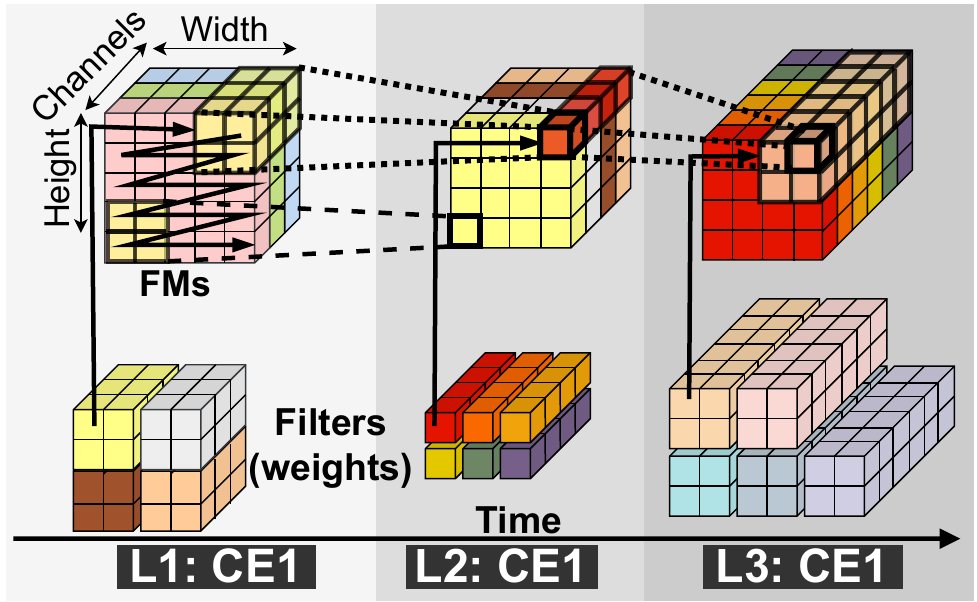}
         \caption{Single-CE sequential processing}
         \label{subfig:lbl_processing}
     \end{subfigure}
      \hfill
 \begin{subfigure}[t]{0.39\textwidth}
         \centering
         \captionsetup{justification=centering}
         \includegraphics[width=\textwidth]{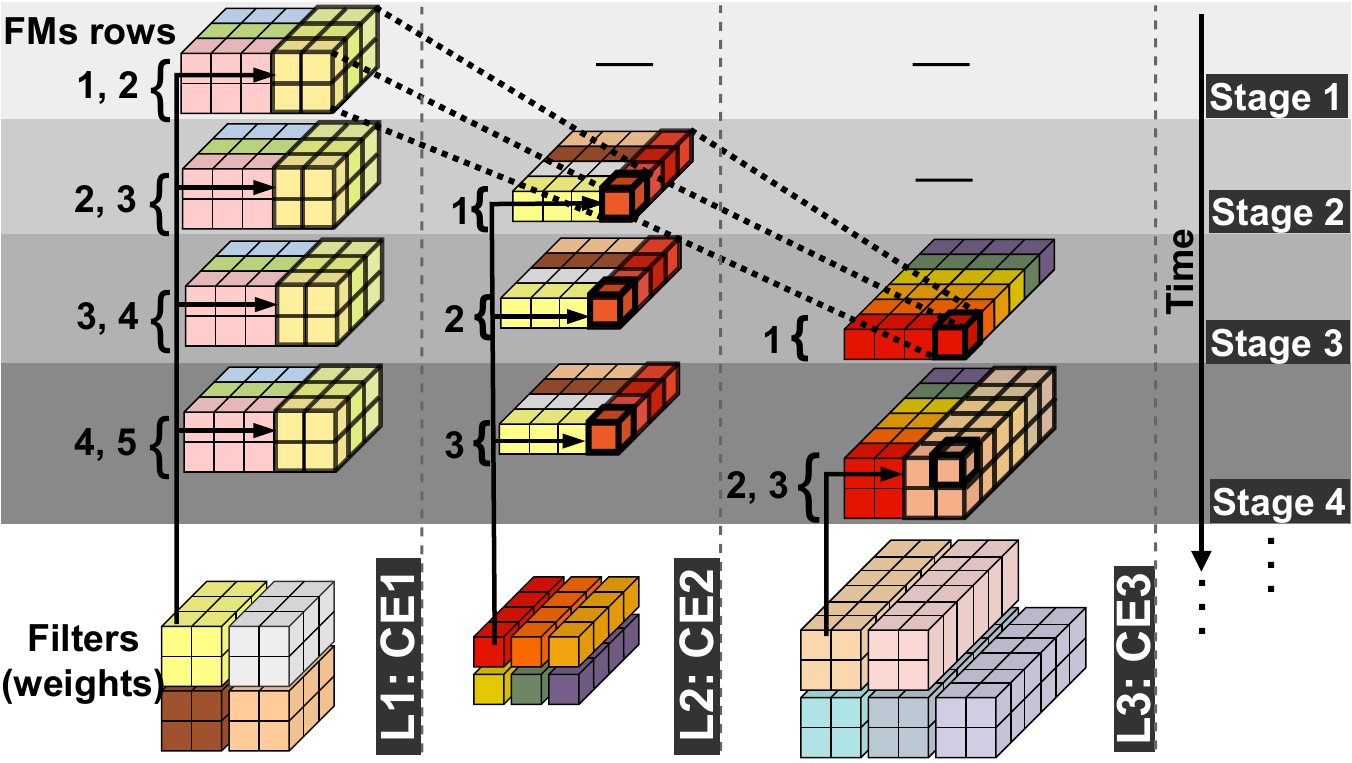}
         \caption{Pipelined-CEs concurrent processing}
         \label{subfig:pipe_processing}
     \end{subfigure}
     \hfill
 \begin{subfigure}[t]{0.23\textwidth}
         \centering
         \captionsetup{justification=centering}
         \includegraphics[width=\textwidth]{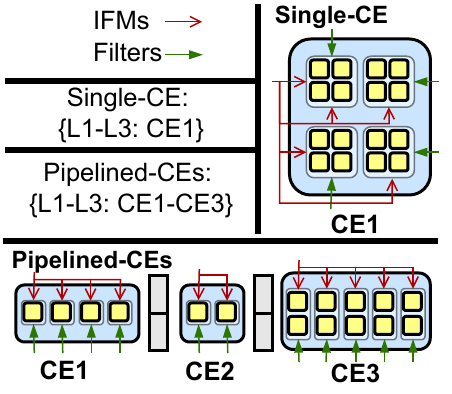}
         \caption{16 PEs distribution examples}
         \label{subfig:sample_engines}
     \end{subfigure}
\caption{\centering Sequential (layer by layer) and pipelined processing of three convolutional layers}  
\label{fig:conv_processing}
%\vspace{-10pt}
\end{figure*} 

This section presents the modeling of the single and pipelined-CEs' blocks. Figure~\ref{subfig:lbl_processing} and~\ref{subfig:pipe_processing} illustrate the differences between processing three convolutional layers using the two blocks. Figure~\ref{subfig:lbl_processing} depicts the single-CE case. In this case, each layer is processed to completion before moving to the next. Figure~\ref{subfig:pipe_processing} depicts the pipelined-CEs case. In this case, the layers are processed concurrently, stage by stage, at tile granularity. The figure shows four out of the six stages.

\subsubsection{Latency and throughput}
\label{subsec:latency_throughput}

In the \emph{single-CE} case, the latency of processing a set of layers is the sum of their latencies, and the throughput is its inverse. 
% The latency of an individual layer, assuming an engine with parallelism P does P MAC operation in one cycle, is the number of MAC operations divided by P plus the extra overheads due to PE underutilization. 
For simplicity, we assume that all layers are compute-bound and that the memory access time is completely hidden by computation time throughout the discussion. In practice, however, we do consider memory access time. Equation~\ref{eq:lbl_latency} describes latency estimation considering PE underutilization, where $layers$ is the number of layers processed by the single-CE, \textit{DD} is the disjoint dimensions of the filters and FMs, \emph{i.e.} the six dimensions corresponding to the six convolution loops (Section~\ref{subsec:ce_parallelism}), and $Par(\textit{CE}_{j}, d)$ is the parallelism on a dimension $d$. The parallelism across all dimensions is upper-bounded by the number of PEs of a CE. To explain the PE underutilization, consider the single-CE shown in Figure~\ref{subfig:sample_engines} with the parallelism of $16$ $(4 \times 2 \times 2)$. This CE processes weights from $4$ filters and a $2 \times 2$ IFM tile in a cycle. When this CE is used to process L2 from Figure~\ref{subfig:lbl_processing}, which has $6$ filters, its PEs are fully utilized when processing the first $4$ filters and half utilized when processing the remaining $2$. 
 %Underutilization results in higher latency and lower throughput than achieved with full utilization.
The parallelism dimensions could be configured differently to avoid underutilization in L2, but such arrangements would result in underutilization in the other layers. Generally, the more diverse the layers processed by a single-CE are, the harder it is to avoid PE underutilization.

\begin{equation}
%\vspace{-2pt}
\setlength\abovedisplayskip{0pt}
\label{eq:lbl_latency}
\begin{split}
Latency =& \sum_{i=1}^{layers}{Lat(L_i, \textit{CE}_j)}\\
Lat(L_i, \textit{CE}_j) =& \prod_{d}^{\textit{DD}_i}\ceil{\lvert d \rvert / Par(\textit{CE}_{j}, d)}\\
where~& \prod_{d}^{\textit{DD}_i}Par(\textit{CE}_{j}, d) \leq \textit{PE}s(\textit{CE}_j)
\end{split}
%\vspace{-2pt}
\end{equation}

The literature defines the latency of \emph{pipelined-CEs} in two ways. The first is the time to process a single input, e.g., an image. The second is the time to process a batch of inputs divided by the number of inputs. Here, we adopt the first definition because batching is not always an option and because the first definition is the one that matters in latency-critical applications. Equation~\ref{eq:pipelined_latency} shows latency estimation of pipelined-CEs, where $PipeStages$ is the number of stages to complete processing an input, \textit{CEs} is the number of CEs in the pipeline, $\textit{FMsTile}_{ij}$ is the FMs tile processed at stage $i$ by \textit{CE}$_j$, and $\textit{activeCEs(stage}_i)$ are CEs active at stage $i$. Figure~\ref{subfig:pipe_processing} visualizes stages and CE activity. The interpretation of the equation is that the latency of processing a set of layers is the summation of the pipeline stage latencies, and the stage latency depends on the slowest active CE in that stage.

\begin{equation}
%\vspace{-2pt}
\label{eq:pipelined_latency}
\begin{split}
Latency=& \sum_{i=1}^{PipeStages}{Lat(stage_i)}\\
Lat(stage_i) =& \max_{j=1}^{\textit{CEs}}{\Bigl(Lat(\textit{FMsTile}_{ij}, \textit{CE}_j)\Bigl)}\\
&
where~ \textit{CE}_j \in \textit{activeCEs}(stage_i)
\end{split}
%\vspace{-2pt}
\end{equation}

Equation~\ref{eq:pipelined_throughput} depicts pipelined-CEs throughput estimation where $\textit{CE\_Stages}_i$ denotes the stages in which $\textit{CE}_i$ is active. Pipeline throughput is the inverse of the slowest CE latency~\cite{zhang2018dnnbuilder}. To maximize the throughput of pipelined-CEs, the slowest CE's latency must be minimized. This is done by balancing the pipeline stages, \emph{i.e.} assigning PEs to each CE proportional to its relative workload~\cite{blott2018finn,zhang2018dnnbuilder}.

\begin{equation}
%\vspace{-2pt}
\label{eq:pipelined_throughput}
\begin{split}
Throughput =& \frac{1}{\max_{i=1}^{\textit{CEs}}{\Bigl(Lat(L_i, \textit{CE}_i)\Bigl)}}\\
Lat(L_i, \textit{CE}_i) =& \sum_{j=1}^{ \textit{CE\_Stages}_i}{Lat(\textit{FMsTile}_{ij}, \textit{CE}_i)}
\end{split}
%\vspace{-2pt}
\end{equation}

The pipelined-CEs example in Figure~\ref{subfig:sample_engines} demonstrates how distributing the PEs on multiple CEs could solve the PE underutilization. Each CE in the figure has a parallelism that perfectly divides the corresponding layer filters and FM dimensions. As a result, the pipelined design has an overall higher throughput than the single-CE. However, this comes at the cost of often higher latency. While improving PE utilization results in lower average latency per operation when all pipeline CEs are active, all CEs are not always active from a single-input (image) perspective. This idleness of certain CEs at certain stages (Figure~\ref{subfig:pipe_processing}) increases the overall latency, this effect scales with the number of CEs~\cite{cai2022deepburning, qararyah2024efficient}. In practice, there is a trade-off between underutilization and latency. CNNs have tens to hundreds of layers; hence, solving PE underutilization completely may require large numbers of dedicated CEs, resulting in higher latency. 

\subsubsection{On-chip buffer requirements}

There is a trade-off between on-chip buffer sizes and off-chip memory accesses. This section describes the buffer sizes needed to guarantee minimum off-chip accesses. For generality, we assume weights are stored off-chip at the beginning of a CNN inference. This is because CNNs may have several millions of weights that cannot always be stored fully on-chip. Consequently, \emph{minimum off-chip accesses} in this context refer to one access per weight and zero access per FM element (apart from the mandatory loads and stores of CNN first and last layers FMs). In \emph{single-CE} case, layers are processed one at a time. Hence, the OFMs are produced completely at the end of each layer. But only a portion of layer weights must be on-chip at a time (See Figure~\ref{subfig:lbl_processing}). Equation~\ref{eq:lbl_on_chip_buffer} depicts the estimation of buffer size requirements, where $Sz$ stands for size. The same buffers are reused across layers because layers are processed one at a time. Consequently, the buffering required to achieve the target minimum accesses must accommodate the largest layer FMs, both IFMs and OFMs, plus the largest weights tile. Note that $\textit{FMs}_i$ must account for multiple copies of the FMs in case a layer has residual connections~\cite{he2016deep} (e.g., $L_2$ in Figure~\ref{fig:multi_ce_mappings_detailed}).

\begin{equation}
%\vspace{-2pt}
\label{eq:lbl_on_chip_buffer}
\begin{split}
\textit{BufferSz}  = \max_{i=1}^{layers}{( \textit{FMsSz}_{i})} + \max_{i=1}^{layers}{( \textit{weightsTileSz}_{i})}
\end{split}
%\vspace{-2pt}
\end{equation}

In \textit{pipelined-CEs} case, achieving the defined minimum off-chip accesses requires that all the weights of the pipelined layers be kept on-chip after their first load. Otherwise, as can be inferred from Figure~\ref{subfig:pipe_processing}, each weight needs to be accessed as many times as the pipeline stages in which its CE is active. Regarding the FMs, double buffering is required between the CEs to work concurrently. Equation~\ref{eq:pipe_on_chip_buffer} describes the on-chip buffer requirements of pipelined-CEs to guarantee minimum off-chip accesses. Multiplication by $2$ is because of double buffering. \textit{FMsBufferSz} is determined per layer using Multiple-CE builder heuristics. Smaller buffers mean smaller on-chip memory, but less reuse and more weight movement~\cite {zhang2018dnnbuilder, qararyah2024efficient}.

\begin{equation}
%\vspace{-2pt}
\label{eq:pipe_on_chip_buffer}
\begin{split}
 \textit{BufferSz} = \sum_{i=1}^{ \textit{layers}}{\Bigl( \textit{weightsSz}_{i} + 2 \times \textit{FMsBufferSz}_{i} \Bigl)}
\end{split}
%\vspace{-2pt}
\end{equation}

This section discussed buffer sizes that guarantee the defined minimum off-chip accesses, assuming unlimited on-chip memory. The following section covers the case where these buffers exceed the available on-chip memory.

\subsubsection{Off-chip access}

Equation~\ref{eq:lbl_off_chip} depicts an example of calculating off-chip accesses of a \textit{single-CE} assuming an OS dataflow (Section~\ref{sec:background}), where $\textit{offCh(x)}$ evaluates to $1$ if \textit{x} is stored off-chip and $0$ otherwise. As discussed, we assume that the weights are stored off-chip at the beginning of an inference for generality. Hence, the ideal scenario is when both IFMs and OFMs of a layer fit on-chip, resulting in accesses equal to the size of the weights. \coloredtxt{ The equation also describes two options when the available on-chip memory does not guarantee the ideal scenario. The first option is an output stationary local input stationary. In this option, an IFM element is loaded once from off-chip memory and used to produce all the dependent OFM elements. In this option, the weights need to be loaded multiple times. The second option is output stationary local weight stationary. This option is the opposite of the first, meaning that a weight is loaded once, but an IFM element is loaded multiple times. The adopted option is the one that has fewer memory accesses.} Multiple-CE Builder heuristics identify the buffer sizes that minimize accesses in each option.

\begin{equation}
\label{eq:lbl_off_chip}
\begin{split}
Accesses& = \sum_{i=1}^{layers}{Acc(L_i, \textit{CE}_j)}\\
\textit{Acc(L}_i, \textit{CE}_j)& = argmin \bigg( \textit{OFMsSz}_i \times \textit{offCh(OFMs}_i) + \\
&
\textit{min}\Bigl( \textit{weightsSz}_i\ \times \ceil{\frac{\textit{IFMsSz}_i}{ \textit{IFMsBufferSz}_j }} + \\
&
\textit{IFMsSz}_i , \textit{IFMsSz}_i\ \times
\ceil{\frac{\textit{weightsSz}_i}{ \textit{weightsBufferSz}_j}} + \\
&
\textit{weightsSz}_i \Bigl) \times \textit{offCh(IFMs}_i) + \\
&
\Bigl(1 - \textit{offCh(IFMs}_i) \Bigl) \times \textit{weightsSz}_i \bigg)\\
s.t.\ \textit{weights}&\textit{BufferSz}_j + \textit{IFMsBufferSz}_j + \max_{i=1}^{layers}\bigl(
\textit{OFMsSz}_i \times\\
&(1 - \textit{offCh(OFMs}_i))\bigl) \leq \textit{CE\_BufferSz}_{j}
\end{split}
\end{equation}

Equation~\ref{eq:pipe_off_chip} gives off-chip accesses in the \textit{pipelined-CEs} case where $\textit{offCh(weights}_i, j)$ evaluates to 1 if the weights of the layer are off-chip at the beginning of stage $j$. As the weights are initially off-chip, $\textit{offCh(weights}_i, 1)$ is always 1. In other stages, the weights of a layer are stored on-chip if the space allows. Each weight that is not kept on-chip after the first load must be accessed as many times as the pipeline stages in which its layer is active (see Figure~\ref{subfig:pipe_processing}). In the pipelined-CEs case, the FMs are kept on-chip. This is practically the case when doing tile-grained pipelining, as the buffer sizes are tailored to the available on-chip memory.

\begin{equation}
%\vspace{-2pt}
\label{eq:pipe_off_chip}
\begin{split}
Accesses =\ & \sum_{i=1}^{\textit{CEs}}{\sum_{j=1}^{\textit{CE\_Stages}_i}{\textit{weightsSz}_i \times \textit{offCh(weights}_i, j)}}
\end{split}
%\vspace{-2pt}
\end{equation}

\subsection{Bottom-up modeling: From blocks to full accelerator}
\label{sub_sec:from_b_to_m}

Modeling a multiple-CE accelerator using the models of its building blocks and the interfacing between them is achieved using generalized versions of the equations from the previous section (Section~\ref{sec:blocks_modeling}) or composite equations formed by substituting some into others. The nature of the modifications to the modeling in the previous section depends on \textbf{(1)} whether the building block processes \textit{one or multiple segments} of a CNN, and \textbf{(2)} whether there is an \textit{inter-segment pipelining} across segments (coarse-grained). This dependency is described in the rest of this section at a high level and clarified using \textit{Segmented} architecture as a concrete example.

\subsubsection{Latency and throughput} Latency and throughput are estimated using the same equations whether the building block processes one or multiple segments. However, the final values differ because when a block processes multiple segments, its PEs, parallelism, and hence utilization are optimized for the average case rather than for a unique segment. The latency and throughput of segments with inter-segment pipelining are estimated similarly to a pipelined-CEs block. However, there is a difference in how the latencies of the pipeline stages are estimated. This is because, in inter-segment pipelining, each stage could comprise a set of CEs rather than one CE and because the granularity is a whole input rather than a tile. The difference between fine and coarse-grained pipelining can be inferred by comparing the Segmented with the Hybrid and SegmentedRR architectures in Figure~\ref{fig:multi_ce_mappings_detailed}. To account for these differences, equations~\ref{eq:pipelined_latency} and~\ref{eq:pipelined_throughput} are modified in two ways depending on the types of the coarsely-pipelined blocks. For example, in the Segmented case where the coarse-grained blocks are of single-CE type (Figure~\ref{fig:multi_ce_mappings_detailed}): \textbf{(1)} the stage latency estimation part in Equation~\ref{eq:pipelined_latency} is substituted by Equation~\ref{eq:lbl_latency} because each stage is a single-CE processing a set of layers, \textbf{(2)} in Segmented coarse-grained pipeline, each segment processes a distinct image concurrently. Hence, the condition that keeps track of active CEs in Equation~\ref{eq:pipelined_latency} must have an additional parameter to identify which CEs are processing the input for which throughput and latency are being estimated. When there is no inter-segment pipelining, the total latency is simply the sum of the segment latencies, and the throughput is the inverse of the latency. In both cases, there is an extra inter-segment communication latency.

\subsubsection{On-chip buffer requirements}
\label{subsec:from_b_to_c_on_chip}
When a building block processes multiple segments, the on-chip buffer sizes are estimated considering the worst case, meaning that the buffers must accommodate the largest weight and FM tiles of the layers across these multiple segments. When there is inter-segment pipelining, the interfacing between the segments must apply double-buffering at input granularity between each two segments. Equation~\ref{eq:seg_on_chip_buffer}, which estimates Segmented architecture buffer sizes, clarifies these points. In a Segmented architecture, in principle, a CE could process multiple segments. Hence, $\textit{CE\_BufferSz}_i$ estimation in Equation~\ref{eq:seg_on_chip_buffer} is simply a generalization of Equation~\ref{eq:lbl_on_chip_buffer}, applied across CE segments to account for the worst case. Because Segmented has inter-segment pipelining, there is double-buffering between each two segments ($2 \times \textit{interSegBufferSz}_i$).

\begin{equation}
%\vspace{-2pt}
\label{eq:seg_on_chip_buffer}
\begin{split}
 \textit{BufferSz} =& \sum_{i=1}^{\textit{Segments}}{\left(\textit{CE\_BufferSz}_i + 2 \times \textit{interSegBufferSz}_i\right)}\\
\textit{CE\_BufferSz}_i & = 
\max_{j=1}^{ \textit{CE\_Segments}_i}{\Bigl(\max_{k=1}^{layers_j}{( \textit{FMsSz}_k)}\Bigl)}\ +\\
&
\max_{j=1}^{\textit{CE\_Segments}_i}{\Bigl(\max_{k=1}^{ layers_j}{( \textit{weightsTileSz}_k)}\Bigl)}
\end{split}
%\vspace{-2pt}
\end{equation}

When there is no inter-segment pipelining, only a single buffer is reused across segments, but to guarantee the defined minimum off-chip accesses, that buffer must accommodate the largest inter-segment intermediate results. 

\subsubsection{Off-chip access} Off-chip accesses of a set of segments are the sum of intra-segment accesses plus the possible off-chip accesses on the interfacing between them. While the intra-segment off-chip accesses are computed using the same equations whether the same block processes multiple segments or not, there is an indirect dependency. The accesses depend on the on-chip buffer sizes (see Equations~\ref{eq:lbl_off_chip},~\ref{eq:pipe_off_chip}), which in turn depend on whether multiple segments are processed using the same block as discussed in Section~\ref{subsec:from_b_to_c_on_chip}. The off-chip accesses on the interface between the segments depend on the \textit{inter-segment pipelining}. Inter-segment pipelining double-buffering requires more space; when not available on-chip, the data communicated between the segments must be stored and loaded from off-chip. For example, Equation~\ref{eq:seg_off_chip_access} calculates the Segmented architecture off-chip accesses,  where $\textit{Acc(L}_j, \textit{CE}_i)$ is calculated using Equation~\ref{eq:lbl_off_chip}. Segmented architecture has inter-segment pipelining, and the inter-segment FMs are stored off-chip if the on-chip cannot accommodate them.

\begin{equation}
%\vspace{-2pt}
\label{eq:seg_off_chip_access}
\begin{split}
\textit{Accesses}=& \sum_{i=1}^{\textit{Segments}}
\Bigg( \Bigl(\sum_{j=1}^{Layers_i}  \textit{Acc(L}_j, \textit{CE}_i)\Bigl)
+ 2\ \times \\
&
 \textit{interSegBufferSz}_i \times \textit{offCh}(\textit{interSegBuffer}_i)\Bigg)
\end{split}
%\vspace{-2pt}
\end{equation}

This section used the Segmented architecture as a concrete example. SegmentedRR, Hybrid, or any multiple-CE architecture modeling builds on top of the models of the single-CE and pipelined-CEs similarly, taking into account the impact of single-block processing multiple segments and the existence or absence of inter-segment pipelining.
\section{Evaluation}
\label{sec:evaluation}

\subsection{Experimental Setup}
\label{subsection:experimental_setup}

\subsubsection{Platforms and systems}
\label{subsec:platforms}

Table~\ref{tab:boards} shows the FPGA boards used in our evaluation focusing on PEs (DSPs), on-chip memory (Block RAM), and off-chip bandwidth. The boards represent various resource budgets with PEs ranging from hundreds to thousands and on-chip memory ranging from 2.4 to 16 MiB\footnote{The on-chip memory values are reported in MiB rather than Mb, which is commonly used by AMD}. To evaluate the accuracy of \theacronym, we use Vitis-IDE (2021.2) to implement and evaluate a set of multiple-CE accelerators using high-level synthesis (HLS). \coloredtxt{Vitis synthesis results are commonly used both in the literature and as a part of the design process in the industry}. \theacronym~and the evaluation methodology are implemented in Python and C++. The experiments are conducted on a machine with 16 logical cores, Intel(R) Core(TM) i7-10700 CPU @ 2.90GHz.

\begin{table}
\centering
\caption{Evaluation FPGA boards}
\resizebox{\columnwidth}{!}{
\begin{tabular}{lcccc} \hline
\rowcolor{gray!10}
&                                           ZC706   & VCU108    & VCU110 & ZCU102 \\ \hline
\cellcolor{gray!10} DSPs                    & 900   & 768       & 1800   & 2520   \\    
\cellcolor{gray!10} Block RAM(MiB)           & 2.4   & 7.6       & 4      & 16.6    \\
\cellcolor{gray!10} Off-chip memory BW (GB/s)   &3.2    & 19.2      & 19.2   & 19.2     \\ \hline
\end{tabular}
}
\label{tab:boards}
%\vspace{-10pt}
\end{table}

\begin{table}
\setlength{\tabcolsep}{2pt}
\caption{Evaluated CNN models}
\resizebox{\columnwidth}{!}{
\begin{tabular}{lccccc} \hline
\rowcolor{gray!10}
&\cellcolor{gray!10} ResNet152   & \cellcolor{gray!10} ResNet50    & \cellcolor{gray!10} XCeption     & \cellcolor{gray!10} DenseNet121  & \cellcolor{gray!10} MobileNetV2 \\ \hline
 \cellcolor{gray!10}Abbreviation           & Res152 & Res50                           & XCp                              & Dns121                           & MobV2 \\ 
 \cellcolor{gray!10}Weights (M)             & 60.4   & 25.6                            & 22.9                             & 8.1                              & 3.5   \\
 \cellcolor{gray!10}Conv layers            & 155    & 53                              & 74                               & 120                              & 52         \\ \hline 
\end{tabular}
}
\label{tab:cnns}
%\vspace{-10pt}
\end{table}

\subsubsection{Workloads}
\label{subsec:workloads}

We use a representative set of CNNs with different structures, depths, and parameter (weight) counts. These are ResNet152 and ResNet50~\cite{he2016deep}, XCeption ~\cite{chollet2017xception}, MobileNetV2~\cite{sandler2018mobilenetv2}, and DenseNet~\cite{huang2017densely}. Table~\ref{tab:cnns} shows the relevant characteristics of these CNNs. \coloredtxt{The results presented can be generalized as the evaluated CNNs share core blocks with many recent CNNs. For example, MBConv, the core block of MobileNetV2, is used in EfficientNet~\cite{tan2019efficientnet} and MnasNet~\cite{tan2019mnasnet}. The same applies to the Residual blocks of ResNets, and the Extreme Inceptions of XCeption are also used in recent CNNs.}

\subsubsection{Baseline architectures}
\label{subsec:baseline_archs}
To validate \theacronym~ and present a set of use cases, we implement representative accelerators of the three multiple-CE architectures (Section~\ref{subsec:back_multi_ce}). The Segmented implementation is based on Shen \textit{et al.}~\cite{shen2017maximizing}. The SegmentedRR tiling and buffer designs and granularity are based on Wei \textit{et al.}~\cite{wei2018tgpa}, and the engine design is based on Ma \textit{et al.}~\cite{ma2018optimizing}. The implementation of Hybrid architectures is based on Qararyah \textit{et al.}~\cite{qararyah2024efficient}. The experiments use 10 different CE counts in each architecture, starting from 2 CEs (the smallest possible CE count in a multiple-CE accelerator) to 11 CEs. These CE counts cover all possibilities in the baseline architectures. However, \theacronym~does not impose a limitation on the number of CEs. The number of PEs in a CE depends on the PE budget and is proportional to the CE workload. For example, in a multiple-CE accelerator with $2$ CEs on VCU108, the $768$ DSPs are distributed on the $2$ CEs, and in an accelerator with $11$ CEs, the same $768$ DSPs are distributed on the $11$ CEs.

\subsection{\theacronym~validation}
\label{subsec:validation}

\begin{table*}
\centering
\caption{\theacronym~evaluation accuracy on VCU108: summary of 150 experiments (3 architectures $\times$ 10 CE counts $\times$ 5 CNNs)}
%\resizebox{\textwidth}{!}{
\begin{tabular}{lccc|ccc|ccc} \hline
\rowcolor{gray!10}
&\multicolumn{3}{c|}{\cellcolor{gray!10}Max} & \multicolumn{3}{c|}{\cellcolor{gray!10}Min}     & \multicolumn{3}{c}{\cellcolor{gray!10}Average}
 \\
 \rowcolor{gray!10}
 &Segmented&SegmentedRR&Hybrid&Segmented&SegmentedRR& Hybrid&Segmented&SegmentedRR& Hybrid\\ \hline
\cellcolor{gray!10}On-chip buffers&99.4\%&99.7\%&99.8\%&84.2\%&91.2\%&84.6\%& 93.1\%& 97.4\%& 95.4\%\\   
\cellcolor{gray!10}Latency&98.8\%&99.0\%&99.6\%&87.2\%&80.7\%&85.0\%&92.8\%&93.3\%&92.5\%\\
\cellcolor{gray!10}Throughput&99.6\%&100\%&99.6\%&86.4\%&83.4\%&85.0\%&93.9\%&95.1\%& 92.5\%\\
\cellcolor{gray!10}Off-chip accesses&100\%&100\%&100\%&100\%&100\%&100\%&100\%&100\%&100\%\\
 \hline
\end{tabular}
%}
\label{tab:validation}
%\vspace{-10pt}
\end{table*}

Table~\ref{tab:validation} summarizes $150$ experiments validating \theacronym~estimation accuracy. The accuracy of the estimation is calculated using Equation~\ref{eq:accuracy}. \theacronym~off-chip accesses calculations are exact since the accesses are deterministic and independent of the optimizations of the synthesis. \theacronym~accuracy ranges from $80.7\%$ to $100\%$. The average accuracy values are $> 90\%$ for the three architectures. Moreover, the average accuracies of the three architectures are close to each other; the differences among them are $<5\%$. There are differences between latency and throughput accuracies in most cases because in multiple-CE accelerators, throughput is not necessarily the inverse of end-to-end latency (Section~\ref{subsec:latency_throughput}). \coloredtxt{\theacronym~correctly predicted the best architecture among the three in 139 of the 150 experiments when considering on-chip buffer sizes, and in all experiments when considering latency, throughput, and accesses. Performance differences between architectures vary greatly and are sensitive to CE count. For example, considering ResNet50, SegmentedRR has the lowest latency of $73\%$ normalized to the latency of Hybrid and $58\%$ normalized to the latency of Segmented when using 2 CEs. In contrast, Hybrid has the lowest latency of $20\%$ normalized to the latency of Segmented and $96\%$ normalized to the latency of SegmentedRR when using 11 CEs.
}

\begin{equation}
\label{eq:accuracy}
\textit{Accuracy} = 100 \times \Bigl( 1 - \frac{\lvert \textit{synthesis result} - \textit{estimated} \rvert}{\textit{synthesis result}} \Bigl) \%
\end{equation}

\subsection{{\bf Use Case 1}: End-to-end evaluation}
\label{subsec:best_multi_ce}

\begin{table}[!htbp]
\setlength{\tabcolsep}{1pt}
\caption{\centering Multiple-CE accelerators that achieve best results (\fcolorbox{white}{seg_color!70}{\rule{0pt}{2pt}\rule{2pt}{0pt}} Segmented
\fcolorbox{white}{segrr_color!70}{\rule{0pt}{2pt}\rule{2pt}{0pt}} SegmentedRR
\fcolorbox{white}{hyb_color!70}{\rule{0pt}{2pt}\rule{2pt}{0pt}} Hybrid) and their CE-counts. Multiple colored cells in the same column for the same metric indicate a tie. We consider results within a $10\%$ difference as a tie to account for estimation errors.}
\resizebox{\columnwidth}{!}{
\begin{tabular}{|l|c|c|c|c|c|c|c|c|c|c|c|c|c|c|c|c|c|c|c|c|c|c|c|c|} \hhline{~-----------------------}

\multicolumn{1}{c|}{} &\multicolumn{5}{c|}{\cellcolor{gray!10}zc706} &  &\multicolumn{5}{c|}{\cellcolor{gray!10}vcu108} &  &\multicolumn{5}{c|}{\cellcolor{gray!10}vcu110} &  &\multicolumn{5}{c|}{\cellcolor{gray!10}zcu102}\\ \hhline{~-----------------------}
\multicolumn{1}{c|}{} & \cellcolor{gray!10} \rotatebox[origin=c]{90}{Res152} & \cellcolor{gray!10} \rotatebox[origin=c]{90}{Res50} & \cellcolor{gray!10} \rotatebox[origin=c]{90}{XCp} & \cellcolor{gray!10} \rotatebox[origin=c]{90}{Dns121} & \cellcolor{gray!10} \rotatebox[origin=c]{90}{MobV2} &  & \cellcolor{gray!10} \rotatebox[origin=c]{90}{Res152} & \cellcolor{gray!10} \rotatebox[origin=c]{90}{Res50} & \cellcolor{gray!10} \rotatebox[origin=c]{90}{XCp} & \cellcolor{gray!10} \rotatebox[origin=c]{90}{Dns121} & \cellcolor{gray!10} \rotatebox[origin=c]{90}{MobV2} &  & \cellcolor{gray!10} \rotatebox[origin=c]{90}{Res152} & \cellcolor{gray!10} \rotatebox[origin=c]{90}{Res50} & \cellcolor{gray!10} \rotatebox[origin=c]{90}{XCp} & \cellcolor{gray!10} \rotatebox[origin=c]{90}{Dns121} & \cellcolor{gray!10} \rotatebox[origin=c]{90}{MobV2} &  & \cellcolor{gray!10} \rotatebox[origin=c]{90}{Res152} & \cellcolor{gray!10} \rotatebox[origin=c]{90}{Res50} & \cellcolor{gray!10} \rotatebox[origin=c]{90}{XCp} & \cellcolor{gray!10} \rotatebox[origin=c]{90}{Dns121} & \cellcolor{gray!10} \rotatebox[origin=c]{90}{MobV2}\\ \hline
\cellcolor{gray!10} &  &\cellcolor{seg_color!70}2 &\cellcolor{seg_color!70}2 &  &  &  &  &  &\cellcolor{seg_color!70}2 &  &  &  &  &  &  &  &\cellcolor{seg_color!70}2 &  &  &  &  &  & \\ 
\cellcolor{gray!10} &\cellcolor{segrr_color!70}2 &  &  &\cellcolor{segrr_color!70}3 &\cellcolor{segrr_color!70}3 &  &\cellcolor{segrr_color!70}2 &\cellcolor{segrr_color!70}2 &  &\cellcolor{segrr_color!70}3 &\cellcolor{segrr_color!70}3 &  &\cellcolor{segrr_color!70}2 &\cellcolor{segrr_color!70}3 &  &\cellcolor{segrr_color!70}3 &\cellcolor{segrr_color!70}3 &  &\cellcolor{segrr_color!70}2 &\cellcolor{segrr_color!70}2 &\cellcolor{segrr_color!70}2 &\cellcolor{segrr_color!70}4 & \\ 
\cellcolor{gray!10}\multirow{-3}{*}{\cellcolor{gray!10}Latency} &  &  &\cellcolor{hyb_color!70}4 &  &\cellcolor{hyb_color!70}2 &  &  &  &\cellcolor{hyb_color!70}4 &  &\cellcolor{hyb_color!70}3 &  &  &  &\cellcolor{hyb_color!70}4 &  &\cellcolor{hyb_color!70}7 &  &  &  &\cellcolor{hyb_color!70}4 &  &\cellcolor{hyb_color!70}7\\ 
\hline 
\cellcolor{gray!10} &\cellcolor{seg_color!70}3 &\cellcolor{seg_color!70}2 &\cellcolor{seg_color!70}2 &  &\cellcolor{seg_color!70}8 &  &\cellcolor{seg_color!70}3 &\cellcolor{seg_color!70}2 &\cellcolor{seg_color!70}2 &  &  &  &\cellcolor{seg_color!70}3 &\cellcolor{seg_color!70}6 &  &  &\cellcolor{seg_color!70}9 &  &\cellcolor{seg_color!70}4 &  &\cellcolor{seg_color!70}2 &\cellcolor{seg_color!70}5 &\cellcolor{seg_color!70}11\\ 
\cellcolor{gray!10} &  &  &  &  &  &  &  &  &  &  &  &  &  &  &  &  &  &  &  &  &  &  & \\ 
\cellcolor{gray!10}\multirow{-3}{*}{\cellcolor{gray!10}Throughput} &  &\cellcolor{hyb_color!70}10 &  &\cellcolor{hyb_color!70}11 &\cellcolor{hyb_color!70}2 &  &  &\cellcolor{hyb_color!70}6 &  &\cellcolor{hyb_color!70}10 &\cellcolor{hyb_color!70}3 &  &  &\cellcolor{hyb_color!70}11 &\cellcolor{hyb_color!70}2 &\cellcolor{hyb_color!70}11 &  &  &  &\cellcolor{hyb_color!70}8 &  &\cellcolor{hyb_color!70}3 &\cellcolor{hyb_color!70}7\\ 
\hline 
\cellcolor{gray!10} &  &  &  &  &  &  &\cellcolor{seg_color!70}2 &\cellcolor{seg_color!70}2 &  &\cellcolor{seg_color!70}2 &\cellcolor{seg_color!70}2 &  &\cellcolor{seg_color!70}2 &\cellcolor{seg_color!70}2 &\cellcolor{seg_color!70}2 &\cellcolor{seg_color!70}2 &\cellcolor{seg_color!70}2 &  &  &  &  &\cellcolor{seg_color!70}2 & \\ 
\cellcolor{gray!10} &  &  &  &\cellcolor{segrr_color!70}2 &\cellcolor{segrr_color!70}2 &  &\cellcolor{segrr_color!70}2 &\cellcolor{segrr_color!70}2 &\cellcolor{segrr_color!70}2 &\cellcolor{segrr_color!70}2 &\cellcolor{segrr_color!70}2 &  &\cellcolor{segrr_color!70}2 &\cellcolor{segrr_color!70}2 &\cellcolor{segrr_color!70}2 &\cellcolor{segrr_color!70}2 &\cellcolor{segrr_color!70}2 &  &  &  &  &\cellcolor{segrr_color!70}2 &\cellcolor{segrr_color!70}2\\ 
\cellcolor{gray!10}\multirow{-3}{*}{\cellcolor{gray!10}Access} &\cellcolor{hyb_color!70}3 &\cellcolor{hyb_color!70}3 &\cellcolor{hyb_color!70}7 &\cellcolor{hyb_color!70}3 &\cellcolor{hyb_color!70}6 &  &\cellcolor{hyb_color!70}3 &\cellcolor{hyb_color!70}3 &\cellcolor{hyb_color!70}7 &\cellcolor{hyb_color!70}3 &\cellcolor{hyb_color!70}5 &  &\cellcolor{hyb_color!70}2 &\cellcolor{hyb_color!70}2 &\cellcolor{hyb_color!70}6 &\cellcolor{hyb_color!70}2 &\cellcolor{hyb_color!70}5 &  &\cellcolor{hyb_color!70}4 &\cellcolor{hyb_color!70}6 &\cellcolor{hyb_color!70}7 &\cellcolor{hyb_color!70}3 &\cellcolor{hyb_color!70}5\\ 
\hline 
\cellcolor{gray!10} &\cellcolor{seg_color!70}2 &\cellcolor{seg_color!70}2 &\cellcolor{seg_color!70}2 &  &  &  &  &\cellcolor{seg_color!70}2 &  &  &  &  &\cellcolor{seg_color!70}2 &  &  &  &  &  &\cellcolor{seg_color!70}2 &\cellcolor{seg_color!70}2 &\cellcolor{seg_color!70}2 &  & \\ 
\cellcolor{gray!10} &  &  &  &\cellcolor{segrr_color!70}2 &\cellcolor{segrr_color!70}3 &  &  &  &  &\cellcolor{segrr_color!70}2 &\cellcolor{segrr_color!70}3 &  &  &  &  &\cellcolor{segrr_color!70}2 &\cellcolor{segrr_color!70}3 &  &  &  &  &\cellcolor{segrr_color!70}2 &\cellcolor{segrr_color!70}3\\ 
\cellcolor{gray!10}\multirow{-3}{*}{\cellcolor{gray!10}Buffers} &  &\cellcolor{hyb_color!70}2 &\cellcolor{hyb_color!70}6 &  &\cellcolor{hyb_color!70}11 &  &\cellcolor{hyb_color!70}2 &\cellcolor{hyb_color!70}2 &\cellcolor{hyb_color!70}7 &  &\cellcolor{hyb_color!70}11 &  &\cellcolor{hyb_color!70}3 &\cellcolor{hyb_color!70}3 &\cellcolor{hyb_color!70}7 &  &\cellcolor{hyb_color!70}11 &  &  &\cellcolor{hyb_color!70}2 &\cellcolor{hyb_color!70}6 &  &\cellcolor{hyb_color!70}11\\ 
\hline 

% \multicolumn{12}{c}{\fcolorbox{white}{seg_color!70}{\rule{0pt}{2pt}\rule{2pt}{0pt}} Segmented} 
% {\fcolorbox{white}{segrr_color!70}{\rule{0pt}{2pt}\rule{2pt}{0pt}} SegmentedRR}
% {\fcolorbox{white}{hyb_color!70}{\rule{0pt}{2pt}\rule{2pt}{0pt}} Hybrid} 

\end{tabular}
}
\label{tab:bests}
%\vspace{-10pt}
\end{table}

This use case discusses \theacronym~end-to-end evaluation of the baseline multiple-CE architectures. Table~\ref{tab:bests} depicts a summary of this evaluation. It shows \theacronym~selection of the accelerators that achieve the best results considering different metrics, CNNs, and FPGA boards. Four insights can be identified by examining Table~\ref{tab:bests}. \textit{First}, in $80\%$ of the cases, no single architecture provides the best results in the four metrics, even for the same CNN model and board. This is shown in the table as $16$ out of $20$ columns do not have four cells of the same color. The $4$ cases where Hybrid is best in all metrics belong to MobV2 and XCp, these CNNs have various convolution types. This is expected as Hybrid architecture was proposed to handle such CNNs~\cite{qararyah2024efficient}. \textit{Second}, even when a single architecture yields the best results in all metrics, no single instance of that architecture is always the best. For example, for MobV2 on ZC706 (fifth column), the Hybrid accelerator of $2$ CEs has the best latency and throughput, the one with $6$ CEs has the minimum off-chip accesses, and the one with $11$ CEs has the minimum buffer requirements. \textit{Third}, some metrics are uniquely dominated by one of the architectures. For example, SegmentedRR achieves the best latency in $15$ out of $20$ cases. This is because SegmentedRR architectures were proposed to minimize latency~\cite{venieris2017latency, wei2018tgpa}. Hybrid has the lowest buffer requirements in $14$ out of $20$ cases. \textit{Fourth}, the Hybrid always achieves the minimum off-chip accesses as this was one of its design objectives~\cite{qararyah2024efficient}. The other two architectures catch up on VCU108 and VCU110, as these boards have relatively large on-chip memories that accommodate big enough buffers required to achieve the minimum off-chip access (Table~\ref{tab:boards}). \\
\textit{In summary}, multiple-CE accelerators' different CE arrangements affect their performance and efficiency, as discussed in detail in sections~\ref{sec:blocks_modeling} and~\ref{sub_sec:from_b_to_m}. This highlights the importance of exploring different CE arrangement possibilities and identifying the best ones considering the CNN models, hardware resources, and the metric of interest.

\subsection{{\bf Use Case 2}: Fine-grained evaluation}
\label{subsec:perf_res_tradeoff}

\begin{figure}[!htbp]
%\vspace{-10pt}
 \centering
 \captionsetup{justification=centering}
 \includegraphics[]{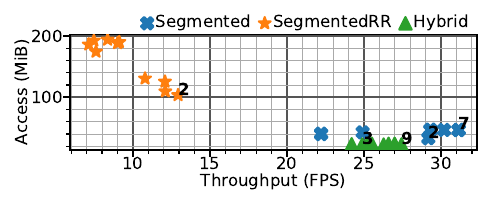}
 %\vspace{-10pt}
 \caption{Throughput vs. off-chip memory accesses of ResNet50 on ZC706 using 10 accelerator instances per architecture with 2-11 CEs. The numbers indicate the CE counts of the accelerators with the highest throughput or minimum accesses of each architecture.}
 %\vspace{-4pt}
 \label{fig:throughput_access}
\end{figure}

\begin{figure}[!htbp]
%\vspace{-6pt}
 \centering
 \captionsetup{justification=centering}
 \begin{subfigure}[t]{\columnwidth}
         \centering
         \captionsetup{justification=centering}
         \includegraphics[]{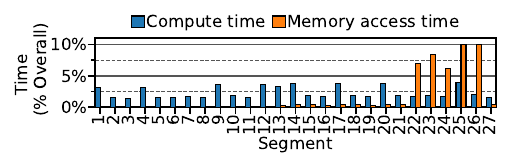}
         %\vspace{-5pt}
         \caption{}
         \label{subfig:segrr_resnet50}
     \end{subfigure}
     \hfill
      \begin{subfigure}[t]{\columnwidth}
         \centering
         \captionsetup{justification=centering}
         \includegraphics[]{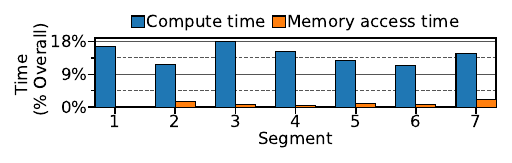}
         %\vspace{-5pt}
         \caption{}
         \label{subfig:seg_resnet50}
     \end{subfigure}
 \caption{Segments compute and memory access time normalized to the overall execution time of \textbf{(a)} SegmentedRR with 2 CEs, \textbf{(b)} Segmented with 7 CEs using ResNet50 on ZC706}
 \label{fig:segrr_resnet50}
 %\vspace{-4pt}
\end{figure}

This use case demonstrates \theacronym~\textit{fine-grained} evaluation to identify the performance bottlenecks of multiple-CE architectures and guide optimizations that alleviate these bottlenecks. Figure~\ref{fig:throughput_access} depicts \theacronym~estimations of throughput and off-chip accesses of the three baseline architectures. As can be seen, both Hybrid and Segmented have relatively good results in both metrics. SegmentedRR instances, however, have considerably more off-chip memory accesses that form a bottleneck. Figure~\ref{subfig:segrr_resnet50} shows \theacronym~breakdown of compute and memory access time of SegmentedRR with 2 CEs, which has the highest throughput among SegmentedRR in Figure~\ref{fig:throughput_access}. In segments $22-26$, the memory access time is the bottleneck. In $29\%$ of the overall execution time, CEs are idle, waiting for data. The Segmented and Hybrid, by contrast, have no such bottlenecks. For example, the breakdown of compute and memory access time of Segmented with 7 CEs, which has the highest throughput among Segmented, is shown in Figure~\ref{subfig:seg_resnet50}.

\begin{figure}[!htbp]
%\vspace{-6pt}
 \centering
 \captionsetup{justification=centering}
 \includegraphics[width=\columnwidth]{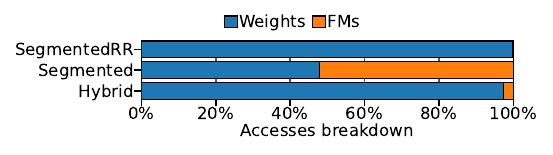}
 %\vspace{-10pt}
 \caption{Off-chip memory accesses breakdown of SegmentedRR with 2 CEs, Segmented with 7 CEs, and Hybrid with 9 CEs using ResNet50 on ZC706.}
 \label{fig:seg_segrr_res50_zc706}
 %\vspace{-10pt}
\end{figure}

When considering alleviating the off-chip access bottleneck, data compression is a candidate optimization~\cite{chen2019eyeriss}. However, compression has its overhead. \theacronym~fine-grained evaluation helps identify the segments that form bottlenecks and apply compression only to these segments' layers to ensure minimum overheads. For example, Figure~\ref{subfig:segrr_resnet50} suggests that compression would be beneficial only for the layers of segments $22-26$. Moreover, it is crucial to identify which data dominates the accesses. For example, Figure~\ref{fig:seg_segrr_res50_zc706} shows \theacronym~breakdown of the accesses of the accelerator instances with the highest throughput in each of the three architectures shown in Figure~\ref{fig:throughput_access}. Figure~\ref{fig:seg_segrr_res50_zc706} suggests that while in SegmentedRR and Hybrid cases, compressing the weights would have a considerable impact on the accesses, compressing FMs would be a pure overhead.

\subsection{{\bf Use Case 3}: Guiding design space exploration}

\begin{figure}[!htbp]
%\vspace{-6pt}
 \centering
 \captionsetup{justification=centering}
 \includegraphics[]{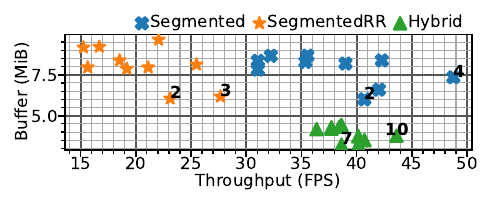}
 %\vspace{-10pt}
 \caption{Throughput vs. on-chip buffer using XCp on VCU110 using 10 accelerator instances per architecture with 2-11 CEs. The numbers indicate the CE counts of the accelerators with the highest throughput or minimum buffer requirements of each architecture.}
 %\vspace{-4pt}
 \label{fig:custom_arch_no_dse}
\end{figure}

This use case demonstrates how  \theacronym~fast evaluation enables a systematic multiple-CE accelerator design approach based on identifying the performance bottlenecks and exploring the space of design points that alleviate these bottlenecks. This use case aims to identify the architecture of a multiple-CE accelerator that maximizes throughput while minimizing on-chip memory usage. The CNN and board used are XCp and VCU110. Figure~\ref{fig:custom_arch_no_dse} shows the trade-off between throughput and on-chip buffer requirements using XCp on VCU110.

One way to derive multiple-CE architectures with better throughput-buffer trade-offs is to take the most promising architectures in each metric as starting points, identify their bottlenecks using \theacronym~fine-grained evaluation, and explore architectures that mitigate these bottlenecks. The most promising architectures in Figure~\ref{fig:custom_arch_no_dse} are those located in the bottom right region. For example, the Segmented accelerator with $4$ CEs has the highest throughput, and the Hybrid with $7$ CEs has the lowest buffer requirements.

\begin{figure}[htbp!]
%\vspace{-6pt}
\centering
\begin{subfigure}[t]{0.5\columnwidth}
     \centering
     \captionsetup{justification=centering}
     \includegraphics[width=\textwidth]{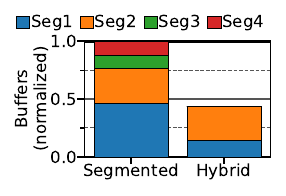}
     \caption{Per-segment buffers normalized to Segmented total buffer size}
     \label{subfig:segmetend_hybrid_buffers}
 \end{subfigure}
 \hfill
 \begin{subfigure}[t]{0.48\columnwidth}
     \centering
     \captionsetup{justification=centering}
     \includegraphics[width=\textwidth]{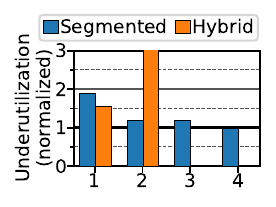}
     \caption{Per-segment PE underutilization normalized to the minimum underutilization}
     \label{subfig:segmetend_hybrid_underutilization}
 \end{subfigure}
\caption{\centering Buffer and PE underutilization of Hybrid with 7 CEs (2 segments) and Segmented with 4 CEs (4 segments)} 
\label{fig:thr_buf_bottlenecks}
%\vspace{-2pt}
\end{figure}

Figure~\ref{fig:thr_buf_bottlenecks} depicts \theacronym~evaluation of the bottlenecks of the two mentioned instances of Segmented and Hybrid architectures. As Figure~\ref{subfig:segmetend_hybrid_buffers} shows, the buffers of the first segments of the Segmented form a bottleneck. The opposite is the case for Hybrid. Regarding throughput, both Hybrid and Segmented have coarse-grained pipelining (Section~\ref{sub_sec:from_b_to_m}). Hence, their throughput is determined by the slowest segment execution time. \theacronym~breakdown shows that the first block is the bottleneck in the case of Segmented, but the last block is the bottleneck in the case of Hybrid. The main reason behind that is the PE underutilization shown in Figures~\ref{subfig:segmetend_hybrid_underutilization}. The higher the PE underutilization is, the slower the block.

\begin{figure}[!htbp]
%\vspace{-10pt}
 \centering
 \captionsetup{justification=centering}
 \includegraphics[]{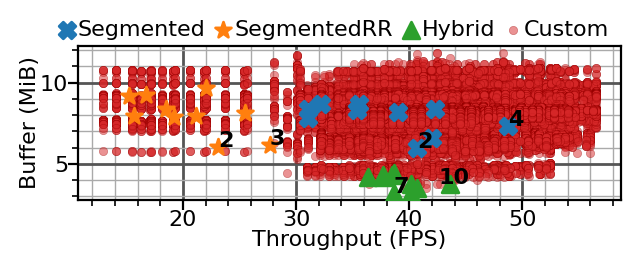}
 %\vspace{-10pt}
 \caption{Throughput vs. on-chip buffer using XCp on VCU110 of a sample of 100000 custom accelerators}
 \label{fig:custom_arch_dse}
 %\vspace{-4pt}
\end{figure}

The observation that the bottlenecks are in the Segmented first block and Hybrid second block hints that a custom architecture that comprises a Hybrid-like first block followed by Segmented-like blocks may improve efficiency. Assuming 10 CE possibilities (2-11 CEs) and given the XCp model structure, the design space of such custom architecture has roughly \emph{97.1 billion} designs. Figure~\ref{fig:custom_arch_dse} shows an evaluation of a random sample of $100000$ out of them. \theacronym~evaluation of this sample took $10.5$ minutes with an average of \emph{6.3 ms} per design. The average synthesis time of a single design on the same machine is roughly an hour, meaning that \theacronym~is in the order of $\emph{100000} \times$ faster. Traditional synthesis-based evaluation of the sample of $100000$ designs would take \emph{years}. Exploring the selected sample resulted in identifying custom accelerators that outperform the state-of-the-art. Some custom accelerators achieve the throughput of the Segmented with 4 CEs while reducing the buffer requirements by up to $48\%$. Custom accelerators with the highest throughput improve throughput by up to $17\%$ compared to Segmented with 4 CEs while reducing buffer requirements by up to $39\%$.
\section{Related work}
\label{sec:related_work}
The literature on DL accelerator design suggests that no fixed architecture is optimal given the diverse workloads and application requirements and that model-aware CNN, and DNN in general, to hardware mappings achieve the best performance and efficiency~\cite{parashar2019timeloop, kwon2019understanding, blott2018finn, boroumand2021google, cai2022deepburning}. The prior art on optimizing CNN to hardware mapping takes two main forms. The first form focuses on intra-layer optimizations, including dataflow and reuse, parallelism strategies, tiling, and exploring custom algorithms like GEMM and FFT~\cite{kao2020gamma, parashar2019timeloop, chen2016eyeriss, kwon2019understanding, ma2018optimizing, lu2017flexflow, lavin2016fast,chetlur2014cudnn}. Timeloop~\cite{parashar2019timeloop} and MAESTRO~\cite{kwon2019understanding} provide frameworks for evaluating and exploring the intra-layer architecture design space. The second form focuses on inter-layer or model-level optimizations, including designing layer-dedicated compute engines (CEs), convolution layer-fusion, and inter-layer pipelining~\cite{blott2018finn, venieris2016fpgaconvnet, cai2021optimus, gao2019tangram, yang2023isosceles, qararyah2024fusing, zhang2018dnnbuilder}.

FPGAs are a common target for inter-layer optimizations due to their reconfigurability, which enables arranging the resources into a custom number of dedicated CEs. When designing a multiple-CE accelerator, there are numerous CE arrangement possibilities. An obvious arrangement is a set of layer-specific and fully pipelined CEs where the number of CEs is equal to the number of CNN layers~\cite{blott2018finn, venieris2016fpgaconvnet,zhang2018dnnbuilder}. More scalable, resource and latency-aware arrangements have an adjustable number of CEs, which is determined by both the CNN structure and the hardware budget~\cite{alwani2016fused, zhang2020dnnexplorer, nguyen2020layer, qararyah2024efficient, shen2016overcoming, shen2017maximizing, venieris2017latency, wei2018tgpa, cai2022deepburning}.\\
The prior art multiple-CE accelerators follow a set of fixed design templates and are not based on systematic design space exploration. This paper proposes a multiple-CE accelerator analytical cost model and a fast evaluation methodology based on this model. Fast evaluation of multiple-CE architectures is key to a systematic design approach.

%FINN is a representative example of this approach~\cite{blott2018finn}, it offers a framework that generates a model-specific streaming-dataflow accelerator where each layer has its own dedicated engine. 
\section{Conclusion}
\label{sec:conclusion}
Multiple-CE accelerators are more adaptable to various CNN model structures and application performance metrics than generic accelerators. However, existing multiple-CE accelerators have fixed architectures and do not explore the impact of CE arrangements on performance and efficiency. This paper proposed \underline{M}ultiple-\underline{C}E accelerator analytical \underline{C}ost \underline{M}odel (\theacronym), and a fast \theacronym-based evaluation methodology. \theacronym~is in the order of $100000 \times$ faster than traditional synthesis-based evaluation and has an average accuracy of $> 90\%$. \theacronym~helps to identify the best-performing among state-of-the-art multiple-CE accelerators given various metrics, CNNs, and resource budgets. Moreover, \theacronym~fast evaluation permits identifying performance bottlenecks and exploring the vast space of CE arrangements, opening the door to a more systematic approach to multiple-CE accelerator design.

% \textcolor{red}{We plan to build a multiple-CE accelerator design framework that utilizes \theacronym~fast evaluation to automate the DSE and the identification of the best-performing multiple-CE accelerators.}

\section*{Acknowledgement}
\coloredtxt{This work was supported by the VEDLIoT project, which received funding from the European Union's Horizon 2020 research and innovation program under grant agreement No 957197. This work was also partly supported by the Swedish Foundation for Strategic Research (contract number CHI19-0048) under the PRIDE project and the European High-Performance Computing Joint Undertaking (JU) under Framework Partnership Agreement No 800928 and Specific Grant Agreement No 101036168 (EPI SGA2).
The JU receives support from the European Union’s Horizon 2020 research and innovation program and from Croatia, France, Germany, Greece, Italy, Netherlands, Portugal, Spain, Sweden, and Switzerland.
}

\bibliographystyle{IEEEtranS}
\bstctlcite{IEEEexample:BSTcontrol}
\bibliography{references}
\end{document}